\documentclass[12pt]{article}
\newcommand{\be}{\begin{equation}}
\newcommand{\ee}{\end{equation}}
\newcommand{\ben}{$$}
\newcommand{\een}{$$}
\newcommand{\bea}{\begin{eqnarray}}
\newcommand{\eea}{\end{eqnarray}}
\newcommand{\bean}{\begin{eqnarray*}}
\newcommand{\eean}{\end{eqnarray*}}
\newcommand{\e}{{\rm e}}

\newcommand{\px}{\langle \hat{\varphi}^2_\ell \rangle (x)}
\renewcommand{\P}{\langle \hat{\Phi}^2 \rangle}
\newcommand{\p}{\langle \hat{\varphi}^2_\ell \rangle}

\newcommand{\fourd}{four-dimensional }

\newcommand{\twod}{two-dimensional }

\newcommand{\n}[1]{\label{#1}}

\newcommand{\ind}[1]{\mbox{\tiny{#1}}}

\usepackage{latexsym}
\usepackage{fullpage}
\usepackage{graphics}
\usepackage{rotating}
\usepackage{epsfig}
\usepackage{psfig}

\begin{document}


\title{{\hfill {\small Alberta-Thy-02-00 } } \vspace*{2cm} \\
The Dimensional-Reduction Anomaly in Spherically Symmetric Spacetimes}
\author{\\
P. Sutton\thanks{e-mail: psutton@phys.ualberta.ca} \\ 
{\em \small Theoretical Physics Institute, Department of Physics,} \\
{\em \small University of Alberta, Edmonton, Canada T6G 2J1} } 
\date{April 30, 2000}
\maketitle
\noindent


\begin{abstract}
In $D$-dimensional spacetimes which can be foliated by $n$-dimensional 
homogeneous 
subspaces, a quantum field can be decomposed in terms of modes on the 
subspaces, reducing the system to a collection of $(D-n)$-dimensional 
fields.  This allows one to write bare  $D$-dimensional field quantities 
like the Green function and the effective action as sums of their 
$(D-n)$-dimensional counterparts in the dimensionally reduced theory. 
It has been shown, however, that renormalization breaks this 
relationship between the original and dimensionally reduced theories, 
an effect called the dimensional-reduction anomaly.  We examine the 
dimensional-reduction anomaly for the important case of spherically 
symmetric spaces.
\end{abstract}
\vspace{.3cm}


\section{Introduction}
\label{intro}

Spacetimes with continuous symmetries play an important role in 
quantum field theory for the reason that symmetries often 
allow one to reduce greatly the computational difficulty 
of a given problem, making practical calculations feasible.  
Of particular interest is the separation of variables approach 
to solving partial differential equations in geometries
with a high degree of symmetry.  
For example, expanding a field propagating in a spherically symmetric
geometry in terms of spherical harmonics and substituting 
into the field equation reduces the system from that of 
a single field $\hat{\Phi}(t,r,\theta,\phi)$ in four dimensions 
to a collection of effective \twod fields $\hat{\varphi}_{\ell}(t,r)$, 
one for each spherical harmonic $Y_{\ell m}(\theta,\phi)$.
In principle, after solving the simpler two-dimensional problems,  
one can obtain quantities 
like the stress tensor or the effective action  
for the original four-dimensional field 
theory by summing the corresponding results for 
the two-dimensional field theories over all modes.

In a previous paper \cite{FrSuZe:00}, it was noted that 
separation of variables can break down when applied to quantum 
field theory, so that summing over the dimensionally 
reduced results no longer yields the corresponding 
quantity in four dimensions.  This occurs because in quantum 
field theory, quantities of physical interest, such 
as the effective action, stress energy tensor, and square 
of the field operator, are divergent and must be renormalized.
While the bare field can be dimensionally reduced into the 
sum of lower-dimensional fields, the divergent parts 
which are to be subtracted in four dimensions generally do not 
equal the sum of the corresponding divergent terms from 
the two-dimensional theories.  As a result, one obtains an incorrect 
answer if one calculates a renormalized quantity  
in four dimensions by summing over modes of 
the corresponding renormalized quantities in two dimensions.  This failure  
of dimensional reduction when applied to quantum field theories
is called the {\bf dimensional-reduction anomaly}.     

In this paper we calculate the dimensional-reduction anomaly 
which occurs when a scalar field propagating in a 
spherically symmetric four-dimensional spacetime is decomposed into 
spherical harmonics and treated as a collection of two-dimensional  
fields.  This case may be of particular importance 
to recent attempts to calculate the stress tensor and 
Hawking radiation in black-hole spacetimes using 
two-dimensional dilaton gravity models \cite{MuWiZe:94,BaFa:99}. 
Since four-dimensional renormalized quantities will not in 
general equal the sum of the corresponding two-dimensional 
quantities, it may well be necessary to take into 
account the contribution of the dimensional-reduction 
anomaly in order to reproduce the correct results 
for four dimensions (see also \cite{NoOd:99}).

We begin in Section~\ref{dimred} with a brief discussion 
of dimensional reduction in spherically symmetric spacetimes.
In Section~\ref{flat} we examine the simple case of flat space, 
both to illustrate the basic idea behind the anomaly and to lay the 
necessary computational groundwork.  
In Sections~\ref{phi_anom} and~\ref{W_anom} 
we extend our calculations to a general 
spherically symmetric four-dimensional space, and calculate the 
dimensional-reduction anomalies in $\langle \hat{\Phi}^2 \rangle$ 
and the effective action.  We conclude with a brief 
discussion of the possible implications of the anomaly.  
We work in Euclidean signature, using dimensionless units where  
$G=c=\hbar=1$ and the sign conventions of \cite{MTW} for the 
definition of the curvature.


\section{Spherical Decompositions}
\label{dimred}

In this section we briefly consider the dimensional reduction of 
a quantum field in a four-dimensional spherically symmetric space, 
and show how it may be reduced to a collection of two-dimensional 
fields.

The line element for such a space may be written as
\be\n{2.1}
ds^2  =  g_{\mu\nu}(X^\tau) dX^\mu dX^\nu  
      =  h_{ab}(x^c) dx^a dx^b 
         + \rho^2 \e^{-2\phi(x^c)} \omega_{ij}(y^k) dy^i dy^j \, ,
\ee
where $X^\alpha=(x^a,y^i)$, 
$h_{ab}$ is an arbitrary two-dimensional metric,  
$\omega_{ij}$ is the metric of a two-sphere, 
$\rho$ is a constant with the dimensions of length, 
and $\phi$ is known as the dilaton.
The radius of a two-sphere of fixed $x^a$ is given by 
$r = \rho \e^{-\phi(x^a)}$.
    
Consider a massive scalar field propagating on the 
space (\ref{2.1}) and obeying the field equation 
\be\n{2.2}
F \, \hat{\Phi}(X) \equiv \left( \Box-m^2-V \right)\hat{\Phi}(X) = 0 \, ,
\ee
where the potential $V$ is also spherically symmetric.
The corresponding Green function is a solution of the equation 
\be\n{2.3}
F \, G(X,X') = -\delta(X,X') \, .
\ee
Knowledge of the Green function for a given quantum state allows
one to calculate other expectation values of interest, such as 
the square of the field operator, $\langle \hat{\Phi}^2 \rangle$,  
and the stress tensor,  $\langle \hat{T}_{\mu\nu} \rangle$.  

Now consider what happens if we decompose 
$\hat{\Phi}$ in terms of spherical harmonics $Y_{\ell m}(y^i)$ as 
follows:
\be\n{2.4}
\hat{\Phi}(X) = \sum_{\ell=0}^\infty \sum_{m=-\ell}^{\ell} 
                \hat{\varphi}_\ell(x^a) \frac{Y_{\ell m}(y^i)}{r} \, .
\ee
Substitution into (\ref{2.2}) shows that  
$\hat{\varphi}_\ell$ behaves as a field
propagating in the \twod space with line element 
\be\n{2.5}
ds^2 = h_{ab}(x^c) dx^a dx^b \, ,
\ee 
and satisfying the field equation 
\be\n{2.6}
{\cal F}_\ell \, \hat{\varphi}_\ell(x) 
\equiv \left( \Delta-m^2-V_\ell \right)\hat{\varphi}_\ell(x) = 0 \, . 
\ee
Here $\Delta$ is the d'Alembertian operator for the \twod 
metric $h_{ab}$, and the induced potential $V_\ell$ is 
given by 
\be\n{2.7}
V_\ell = V + \frac{\ell(\ell+1)}{r^2} - \Delta\phi + (\nabla\phi)^2 \, .
\ee
The corresponding \twod Green functions ${\cal G}_\ell$ 
satisfy
\be\n{2.8}
{\cal F}_\ell \, {\cal G}_\ell(x,x') = -\delta(x,x') \, ,
\ee
and are related to $G$ via 
\be\n{2.9}
G(X,X')
  =  \sum_{\ell=0}^\infty \frac{(2\ell+1)}{4\pi r r'}
     P_\ell(\cos\lambda) {\cal G}_\ell(x,x') \, ,
\ee
where $P_\ell$ is a Legendre polynomial and 
$\lambda$ is the angular separation of $X$, $X'$. 

Since the square of the field operator is given by the 
coincidence limit of the Green function, equation 
(\ref{2.9}) implies that the \fourd  $\P$ can be obtained 
by solving the two-dimensional theory for $\p$: 
\be\n{2.10}
\langle \hat{\Phi}^2 \rangle
  =  \sum_{\ell=0}^\infty \frac{(2\ell+1)}{4\pi r^2}
     \langle \hat{\varphi}^2_\ell \rangle        \, .
\ee
The Green function, however, diverges in the coincidence limit, 
and must be renormalized to yield a finite 
$\langle \hat{\Phi}^2 \rangle$.  Denoting the 
renormalized and divergent parts by 
the subscripts `ren' and `div' respectively, we have 
\bea
\langle \hat{\Phi}^2 \rangle_{\ind{ren}}
  & = &  \lim_{X'\rightarrow X} G_{\ind{ren}}(X,X')  
         = \lim_{X'\rightarrow X} 
         \left[ G(X,X') - G_{\ind{div}}(X,X') \right] \, , \n{2.11} \\
\langle \hat{\varphi}^2_\ell \rangle_{\ind{ren}}
  & = &  \lim_{x'\rightarrow x} {\cal G}_{\ell |\ind{ren}}(x,x') 
         = \lim_{x'\rightarrow x} \left[ {\cal G}_\ell(x,x') 
           - {\cal G}_{\ell | \ind{div}}(x,x') \right] \, . \n{2.12}
\eea
While the bare quantities $G$ and ${\cal G}_\ell$ 
are related by the mode-decomposition 
relation (\ref{2.9}), we shall find that the divergent parts 
$G_{\ind{div}}$ and ${\cal G}_{\ell | \ind{div}}$ 
are not.  As a result, the renormalized theories in 
two and four dimensions are related not by (\ref{2.9}, \ref{2.10}) 
but rather by 
\be\n{2.13}
G_{\ind{ren}}(X,X') 
  =  \sum_{\ell=0}^\infty \frac{(2\ell+1)}{4\pi r r'} 
     P_\ell(\cos\lambda) \left[ 
         {\cal G}_{\ell | \ind{ren}}(x,x') + \Delta{\cal G}_\ell(x,x')
     \right]  \, ,
\ee
\be\n{2.14}
\langle \hat{\Phi}^2 \rangle_{\ind{ren}}(X) 
  =  \sum_{\ell=0}^\infty \frac{(2\ell+1)}{4\pi r^2} \left[
         \langle \hat{\varphi}^2_\ell \rangle_{\ind{ren}}(x) 
         +\Delta\langle \hat{\varphi}^2_\ell \rangle(x)
     \right] \, ,
\ee
where the anomalous terms are easily shown to be   
\bea
\Delta\langle \hat{\varphi}^2_\ell \rangle(x)
 & = &  \lim_{x'\rightarrow x} \Delta{\cal G}_\ell(x,x')  \nonumber \\
 & \equiv &  \lim_{x'\rightarrow x} \left[  \,
            {\cal G}_{\ell | \ind{div}}(x,x')
            -2\pi r r' \!\int_{-1}^1\! d(\cos\lambda) P_\ell(\cos\lambda) 
                G_{\ind{div}}(X,X') 
        \right]  \, . \n{2.15}
\eea
One can show that similar formulae hold for other 
renormalized quantities, such as the effective action $W$ 
and the stress tensor: 
\be\n{2.16}
W_{\ind{ren}} 
  =  \sum_{\ell=0}^\infty (2\ell+1) \left[
         {\cal W}_{\ell | \ind{ren}} + \Delta{\cal W}_\ell
     \right] \, ;
\ee
\be\n{2.17}
\langle \hat{T}_{\mu\nu} \rangle_{\ind{ren}} 
  =  \sum_{\ell=0}^\infty \frac{(2\ell+1)}{4\pi r^2} \left[
         \langle \hat{{\cal T}}_{\mu\nu} \rangle_{\ell | \ind{ren}} 
         +\Delta\langle \hat{{\cal T}}_{\mu\nu} \rangle_\ell
     \right] \, .
\ee
In each case the anomaly is the difference between the 
divergent subtraction terms for the dimensionally reduced theory 
and the mode-decomposed subtraction terms for the original \fourd 
theory.  

Equations (\ref{2.13}--\ref{2.15}) demonstrate that the 
renormalized value of a field quantity is generally not equal 
to the sum of the same renormalized  
quantities for the dimensionally-reduced theory. 
Rather, a quantity like $\langle \hat{\Phi}^2 \rangle$ can 
be obtained from dimensional reduction only if the 
contribution $\langle \hat{\varphi}^2_\ell \rangle$ for each 
mode $\ell$ is modified by an extra anomalous term.
This failure of dimensional reduction under renormalization 
is the dimensional-reduction anomaly.  
The remainder of this paper is devoted to explicit calculations 
of the anomalies  
in $\langle \hat{\Phi}^2 \rangle$ and $W$ for the important case 
of spherically symmetric geometries, described by (\ref{2.1}). 
For further general discussion of the dimensional-reduction anomaly 
and the related multiplicative anomaly,  
the reader is referred to \cite{FrSuZe:00} and  
\cite{ElVaZe:98,Ev:99}.

At this point some conventions on notation are in order.  
We shall need to be able to distinguish quantities 
like Green functions defined in different dimensions.   
`Ordinary' letters such as $G$, $W$ are used for the original 
\fourd theory, while calligraphic letters such as ${\cal G}$, ${\cal W}$ 
refer to dimensionally reduced quantities.  
The anomalous difference between $A$, ${\cal A}$ is denoted $\Delta{\cal A}$.  
All curvatures will be with respect to $h$ unless explicitly 
labelled otherwise; for example, $R = R[h]$ and $\,^4\!R = R[g]$.
As for differential operators, we shall understand $\Box$ to 
represent the d'Alembertian with respect to $g$, while 
$\Delta$ is the d'Alembertian calculated using the metric $h$. 
Single covariant derivatives will be denoted by $\nabla$; 
there will be no need to distinguish the metric used.
For the dilaton $\phi$ we shall understand $\phi_a$, $\phi_{ab}$, etc. to
denote multiple two-dimensional covariant derivatives of $\phi$ 
calculated using the metric $h$.


\section{The Dimensional-Reduction Anomaly in Flat Space}\label{s3}
\setcounter{equation}0
\label{flat}

The simplest example of the dimensional reduction anomaly occurs in the 
spherical decomposition of a scalar field in flat space, and was   
originally considered in \cite{FrSuZe:00}.   
We reproduce here the main formulae, as we shall 
require them for the generalization to curved space, 
and because some of the notation we use is different from 
that of \cite{FrSuZe:00}. 

Let us assume that the potential $V$ vanishes inside the 
region of interest, and is spherically symmetric outside.  
In this case, the Green function for a given state is 
renormalized by subtracting the Green function for the  
Euclidean vacuum.  In four dimensions the latter is  
\be\n{3.1}
G_{\ind{div}}(X,X')={m\over 4\pi^2\sqrt{2\sigma}} K_1(m\sqrt{2\sigma}) \, ,
\ee
where $\sigma$ is one-half the square of the geodesic distance 
between $X$ and $X'$, and $K_1$ is a modified Bessel function.  
In spherical coordinates $X^{\mu}=(t,r,\theta,\eta)$\footnote{We 
denote the azimuthal coordinate by $\eta$ rather than $\phi$ for
obvious reasons.}  
the line element is 
\be\n{3.2}
ds^2 = dt^2 + dr^2 + r^2 \left( d\theta^2 + \sin^2\!\theta d\eta^2 \right)
       \, ,
\ee      
and 
\be\n{3.3}
2 \sigma = 
(t-t')^2 + (r-r')^2 + 2rr'\left(1 - \cos{\lambda} \right)\, ,
\ee
where $\lambda$ is the angle between $X$ and $X'$, given by 
\be\n{3.4}
\cos\lambda = \cos\theta\, \cos\theta'+\sin\theta\, \sin\theta'\,
\cos(\eta -\eta')\, .
\ee
If before renormalizing we first decompose the field $\hat{\Phi}$ into 
spherical harmonics as in (\ref{2.4}), we will be left with an 
effective field $\hat{\varphi}_\ell$ propagating on the two-dimensional  
space with line element $ds^2=dt^2+dr^2$, where 
$t \in (-\infty,\infty)$ and $r \in [0,\infty)$.
In this case $\sigma$ is given 
by $\frac{1}{2}[(t-t')^2 + (r-r')^2]$, and the theory is 
renormalized by subtracting the two-dimensional vacuum Green function   
\be\n{3.5}
{\cal G}_{\ell|\ind{div}}(x,x') 
  =  {1\over 2\pi} K_0(m\sqrt{(t-t')^2+(r-r')^2}) \, .
\ee

To compare the renormalization of the two- and four-dimensional 
theories and establish the existence of the dimensional-reduction 
anomaly, we decompose $G_{\ind{div}}$ into spherical harmonics.  
Defining the mode decomposition by  
\be\n{3.6}
G_{\ind{div}}(X,X') = \sum_{\ell=0}^\infty \frac{(2\ell+1)}{4\pi r r'}
                      P_\ell(\cos\lambda) G_{\ind{div}|\ell}(x,x')
\ee
in accordance with (\ref{2.9}), we have 
\be\n{3.7}
G_{\ind{div}|\ell}(x,x')  =  2\pi r r' \!\int_{-1}^1\! d(\cos\lambda)
                             P_\ell(\cos\lambda) G_{\ind{div}}(X,X') \, .
\ee 
Inserting (\ref{3.1}) into (\ref{3.7}) and using the well-known 
integral representation for $K_\nu$,
\be\n{3.8}
\int_0^\infty\!dx\,x^{-1-\nu} \exp{ \left\{ -x -\frac{\alpha^2}{4x} \right\} }
  =  2 \left( \frac{2}{\alpha} \right)^\nu \! K_\nu(\alpha) \, ,
\ee 
the integral 
\be\n{3.9}
\int_{-1}^1 dz~P_{\ell}(z)~
	\e^{p(1-z)}=(-1)^{\ell}\e^p\sqrt{2\pi\over p}I_{\ell+1/2}(p)\, ,
\ee
where $I_{\ell+1/2}$ is a modified Bessel function 
(see e.g. \cite{PrBrMa:86}, vol.2, eq.2.17.5.2), and the 
representation 
\be\n{3.10}
I_{\ell+1/2}(p)
  =  {1\over\sqrt{2\pi p}}
     \sum_{k=0}^\ell {(\ell+k)!\over k!(\ell-k)!}~{1\over (2 p)^k}
     \left[ (-1)^k \e^p - (-1)^\ell e^{-p} \right] \, ,
\ee
(see, for example, 8.467 of \cite{GrRy:94}),
we obtain 
\bea
G_{\ind{div}|\ell}(x,x')
  & = &  {1\over 2\pi} 
         \sum_{k=0}^{\ell}{(\ell+k)!\over k!(\ell-k)!} \left[
         (-1)^{k}~{[(t-t')^2+(r-r')^2]^{k/2}\over (2m r r' )^k}
         ~K_k(m\sqrt{(t-t')^2+(r-r')^2})   \right. \nonumber \\
  &   &  \left. \mbox{}
         -(-1)^{\ell}
         ~{[(t-t')^2+(r+r')^2]^{k/2}\over (2m r r')^k}
         ~K_k(m\sqrt{ (t-t')^2+(r+r')^2})
         \right]  . \n{3.11}
\eea

Let us compare this result for the mode-decomposed subtraction terms 
from four dimensions with the subtraction term for the two-dimensional  
theory, (\ref{3.5}).  While ${\cal G}_{\ell|\ind{div}}$ is the 
free-field Green function in two dimensions, it is not difficult to 
verify that $G_{\ind{div}|\ell}$ is the Green function for a field 
propagating in the centrifugal barrier 
potential\footnote{See (\ref{2.7}).  It is easy to verify that 
$\Delta\phi-(\nabla\phi)^2=0$ for the line element (\ref{3.2}).} 
\be\n{3.12} 
V_\ell = \frac{\ell(\ell+1)}{r^2}
\ee 
which obeys Dirichlet boundary conditions at $r=0$.  

Naive renormalization in two dimensions requires subtracting  
${\cal G}_{\ell|\ind{div}}$.  We see, however, that $G_{\ind{div}|\ell}$ 
is the quantity that should be subtracted to yield the 
correct results for the renormalized four-dimensional 
theory\footnote{Examples of the correct procedure of renormalizing 
using the mode-decomposed subtraction terms from four dimensions 
in spherically symmetric spaces can be found in \cite{Ca:80}.}. 
If one was to ignore the anomaly and calculate 
$\P_{\ind{ren}}$ using the two-dimensional $\p_{\ind{ren}}$ 
as in (\ref{2.10}), one would obtain incorrect results, 
such as nonvanishing expectation values for the vacuum state.   
Instead, using (\ref{2.14}, \ref{2.15}, \ref{3.5}, \ref{3.11}) 
one finds that the renormalized 
theories in two and four dimensions are related by 
\ben  
\langle \hat{\Phi}^2 \rangle_{\ind{ren}}(X)
  =  \sum_{\ell=0}^\infty \frac{(2\ell+1)}{4\pi r^2} \left[
         \langle \hat{\varphi}^2_\ell \rangle_{\ind{ren}}(x)
         +\Delta\langle \hat{\varphi}^2_\ell \rangle(x)
     \right] \, ,
\een
where the anomaly is 
\bea
\Delta\langle \hat{\varphi}^2_\ell \rangle  
 & = &  \lim_{x'\rightarrow x} \left[ 
           {\cal G}_{\ell|\ind{div}}
            - G_{\ind{div}|\ell} 
        \right]  \nonumber \\
 & = &  \frac{1}{4\pi} 
             \sum_{k=1}^{\ell} {(\ell+k)!\over (\ell-k)!}
             \frac{1}{k} \frac{(-1)^{k+1}}{(mr)^{2k}} 
         +\frac{(-1)^{\ell}}{2\pi} 
             \sum_{k=0}^{\ell}{(\ell+k)!\over k!(\ell-k)!}
         ~\frac{K_k(2 m r)}{(m r)^k} \, . \n{3.13}
\eea    
For example, for the first two modes the anomalies are 
\bea 
\Delta\langle \hat{\varphi}^2_{\ell=0}\rangle
  & = &  \frac{1}{2\pi} K_0(2mr)  \, , \nonumber \\ 
\Delta\langle \hat{\varphi}^2_{\ell=1}\rangle
  & = &  \frac{1}{2\pi} \left[ \frac{1}{(mr)^2} - K_0(2mr) 
             -\frac{2}{(mr)}K_1(2mr) \right] \, .  \nonumber 
\eea   
Note that the anomaly diverges logarithmically 
as $r\rightarrow0$, and vanishes as $r\rightarrow\infty$.


\section{The Dimensional-Reduction Anomaly in  
$\mathbf{\langle \hat{\Phi}^2 \rangle}$}
\setcounter{equation}0
\label{phi_anom}

In the previous section we considered the dimensional-reduction anomaly 
in $\langle \hat{\Phi}^2 \rangle$ arising from the spherical 
decomposition of a scalar field in flat space.
In this section we shall extend those calculations to 
general four-dimensional spherically symmetric spaces.    

Our system consists of a massive scalar field with arbitrary coupling 
to the four-dimensional scalar curvature, described by (\ref{2.2}) with 
\be\n{4.1}
V = \xi\,^4\!R \, . 
\ee
We assume that the spacetime of interest is given by the line element 
(\ref{2.1}), which in standard spherical coordinates $y^i=(\theta,\eta)$
becomes  
\be\n{4.2}
ds^2  =  h_{ab}(x^c) dx^a dx^b + r^2(d\theta^2+\sin^2\!\theta d\eta^2) \, ,
\ee
with $r=\rho\e^{-\phi(x^c)}$.                       

As we saw in Section~\ref{dimred}, under the dimensional 
reduction (\ref{2.4}), the quantum field $\hat{\Phi}$ reduces 
to a collection of effective fields $\hat{\varphi}_\ell$ 
on the \twod space (\ref{2.5}) with metric $h_{ab}$, 
satisfying the field equation (\ref{2.6}) with induced potential 
\be\n{4.4}
V_\ell = \xi\,^4\!R + \frac{\ell(\ell+1)}{r^2} 
         - \Delta\phi + (\nabla\phi)^2 \, .
\ee
We wish to compute the anomaly associated with 
renormalizing this dimensionally reduced theory 
versus (\ref{2.2}).

A standard approach to renormalization in curved space is via 
the heat kernel.  For the system (\ref{2.2}), the 
heat kernel $K(X,X'|s)$ is a solution of the equation 
\be\n{4.5}
F \, K(X,X'|s) = \frac{d\hphantom{s}}{ds} K(X,X'|s)
\ee
with boundary condition $K(X,X'|s=0) = \delta(X,X')$.  Once 
the heat kernel is known for a given state, both the Green function and 
the effective action may be obtained using 
\be\n{4.6}
G(X,X') = \int_0^\infty ds \, K(X,X'|s)  \, , 
\ee
\be\n{4.7}
W = -\frac{1}{2} \int_0^\infty \frac{ds}{s} \int d^4\!X \sqrt{g}\, K(X,X|s)
    \, .
\ee
Analogous formulae hold for the dimensionally reduced theory with operator
${\cal F}_\ell$, heat kernel ${\cal K}_\ell(x,x'|s)$, Green function
${\cal G}_\ell(x,x')$ and effective action ${\cal W}_\ell$.

The advantage of the heat kernel formulation is that the divergences 
in both the Green function and the effective action 
come from the $s\rightarrow0$ limit of the $s$ integral, 
and the small-$s$ behavior of the heat kernel is known for 
arbitrary curved spaces of any dimension.
In particular, in four dimensions, 
\be\n{4.8}
K(X,X'|s)
  =  \frac{D^{\frac{1}{2}}}{(4\pi s)^2}
     \exp{ \left\{ - m^2s - \frac{\sigma}{2s} \right\} }
     \sum_{i=0}^\infty a_i(X,X') s^i  \, .
\ee      
Here again $\sigma = \sigma(X,X')$ is one-half of the square of the
geodesic distance between the points $X$ and $X'$, while 
$D=D(X,X')$ is the Van Vleck determinant,
\be\n{4.9}
D(X,X') = \frac{1}{\sqrt{g(X)}\sqrt{g(X')}}
          \det \left[  - 
              \frac{\partial\hphantom{X^\mu}}{\partial X^\mu}
              \frac{\partial\hphantom{{X'}^\nu}}{\partial {X'}^\nu}  
              \sigma(X,X') 
          \right]    \, .
\ee
The $a_n$ are the Schwinger-DeWitt coefficients for the operator $F$ 
of (\ref{2.2}).  In the coincidence limit $X' \rightarrow X$ the 
first few of these are 
\bea 
a_0^{\Box-\xi\,^4\!R} 
  & = &  1  \, ,  \n{4.10} \\
a_1^{\Box-\xi\,^4\!R} 
  & = &  \left(\frac{1}{6}-\xi\right)\,^4\!R \, , \n{4.11} \\
a_2^{\Box-\xi\,^4\!R} 
  & = &  \frac{1}{180} \left[ 
             \,^4\!R_{\alpha\beta\gamma\delta}\,^4\!R^{\alpha\beta\gamma\delta} 
             -\,^4\!R_{\alpha\beta}\,^4\!R^{\alpha\beta} 
             +\Box \,^4\!R \,
         \right] \nonumber \\
  &   &  \mbox{}
         +\frac{1}{6}\left(\frac{1}{6}-\xi\right) \Box \,^4\!R   
         +\frac{1}{2}\left(\frac{1}{6}-\xi\right)^2(\,^4\!R)^2  \, . 
         \n{4.12}
\eea            
For the two-dimensional operator ${\cal F}_\ell$ of (\ref{2.6}) 
we only need the Schwinger-DeWitt expansion of the heat kernel 
in the coincidence limit.  This is 
\be\n{4.13}
{\cal K}_\ell(x,x|s)
  =  \frac{1}{4\pi s} \exp{ \left\{ - m^2s  \right\} }
     \left[ 1 + s\left(\frac{1}{6}R - V_\ell \right) 
            + \cdots 
     \right]  \, .
\ee 

Considering (\ref{4.6}, \ref{4.7}), it is clear that 
in four dimensions the divergences in $G$ ($W$) arise from the first 
two\footnote{
Comparing to (\ref{3.8}), one sees that the integral
representation used for $G$ in the previous section was just the 
heat kernel representation (\ref{4.6}) with (\ref{4.8}-\ref{4.11}).}  
(three) terms in the Schwinger-DeWitt expansion for $K$, 
while in two dimensions we need consider only the first term 
(first two terms) in ${\cal K}_\ell$.
The anomaly in $\P$ or $W$ can then be calculated by mode-decomposing 
the appropriate terms from $K$, comparing to the heat kernel 
${\cal K}_\ell$ for the dimensionally-reduced theory, and finally 
integrating the difference over $s$ according to (\ref{4.6}) or (\ref{4.7}).
  
Let us begin with the anomaly in $\P$.  
The divergent part of the Green function in four dimensions is given by
\be\n{4.14}
G_{\ind{div}}(X,X') 
  =  \int_0^\infty\!ds \,  
     K_{\ind{div}}(X,X'|s) \, ,
\ee
where $K_{\ind{div}}$ consists of the first two terms of (\ref{4.8}):
\be\n{4.15}
K_{\ind{div}}(X,X'|s) 
  =  \frac{1}{(4\pi s)^2}
     \exp{ \left\{ - m^2s - \frac{\sigma}{2s} \right\} }
     \left[ 
         \Re^{\Box-\xi\,^4\!R}_0(X,X') + s\Re^{\Box-\xi\,^4\!R}_1(X,X') 
     \right]  \, .
\ee         
Here we use the convenient notation 
\be \n{4.16}
\Re^{\Box-\xi\,^4\!R}_n(X,X') \equiv D^{\frac{1}{2}}(X,X')\, 
                            \, a_n^{\Box-\xi\,^4\!R}(X,X') \, .
\ee
In principle, the anomaly in $\P$ is straightforward to 
calculate.  We mode-decompose $K_{\ind{div}}$  
in terms of Legendre polynomials in the usual manner:
\bea
K_{\ind{div}}(X,X'|s) 
  & = &  \sum_{\ell=0}^\infty \, \frac{(2\ell+1)}{4\pi r r'} \, 
         P_\ell(\cos\lambda) \, K_{\ind{div}|\ell}(x,x'|s)  \, ; 
         \n{4.17} \\
K_{\ind{div}|\ell}(x,x'|s)
  & = &  2\pi r r' \int_{-1}^1 d(\cos\lambda) \, P_\ell(\cos\lambda) \,
         K_{\ind{div}}(X,X'|s) \, . \n{4.18}
\eea
The anomaly in $\P$ is then just the coincidence limit of the difference 
between the subtraction terms in two dimensions and those 
mode-decomposed from four dimensions, integrated over $s$: 
\be\n{4.19}
\Delta \px  
  =  \int_0^\infty \!ds\, \left[ 
         {\cal K}_{\ell|\ind{div}}(x,x|s) - K_{\ind{div}|\ell}(x,x|s) 
     \right] \, .
\ee

We encounter a difficulty, however, when we try to perform the 
mode decomposition.
For a general space, $\sigma$ and the $a_n^{\Box-\xi\,^4\!R}$ 
are known only for infinitesimal separations\footnote{ 
In terms of momentum integrals, finite separations correspond
to the low-frequency regime, where the renormalization
terms are not fixed by the divergences in the theory.}  
of $X$ and $X'$,
while evaluation of the mode-decomposition integral (\ref{4.18}) 
requires knowing $\sigma$ and the $a_n^{\Box-\xi\,^4\!R}$ for finite 
separations of $X$, $X'$ on the two-sphere. 
We proceed by determining an approximate $K_{\ind{div}}$ 
for finite separation based on the following criteria:
\begin{enumerate}
\item Our approximate $K_{\ind{div}}$ must reduce to the 
      known value in the flat-space limit.  
\item Our approximate $K_{\ind{div}}$ must respect the 
      periodicity of the two-spheres (i.e., it must be 
      periodic in the angular separation $\lambda$ with 
      period $2\pi$).
\end{enumerate}

In a previous case \cite{FrSuZe:00} in which the mode decompositions 
were performed over noncompact spaces, the following procedure 
was found to work quite well.  We take $X=(x,y)$ and $X'=(x,y')$; 
i.e., we split the points in the $y$-direction only.  Using the 
well-known short-distance expansions obtained 
in \cite{Chri:76,Chri:78}, $\sigma$ and the 
$\Re_n^{\Box-\xi\,^4\!R}$ are expanded in powers of $(y-y')$, which 
is equivalent to expanding in powers of the curvature.  
These expansions are then substituted into 
$K_{\ind{div}}$ and, assuming small curvatures, 
truncated at first order in the curvature for  
$\Delta\langle\hat{\varphi}^2_\ell\rangle$ and at 
second order for $\Delta{\cal W}_\ell$.  
The mode-decomposition integrals in \cite{FrSuZe:00} can 
then be evaluated with relative ease.  

In the present case, the equivalent procedure is to 
expand $\sigma$ and the $\Re_n^{\Box-\xi\,^4\!R}$ in 
powers of $\lambda^2$, which is easily done; see Appendix~\ref{pointsplit}.  
We also take into account the periodicity of the two-spheres by   
converting our expansions in $\lambda^2$ into expansions in 
$(1-\cos\lambda)$.  Defining $z=\cos\lambda$, we have 
\be\n{4.20}
\lambda^2 = 2(1-z)+\frac{1}{3}(1-z)^2+\frac{4}{45}(1-z)^3+\cdots \, .
\ee 
We then substitute (\ref{4.20}) for each $\lambda^2$, truncating at 
the lowest order in $(1-z)$ which will yield the correct flat-space 
limit.  This replacement of $\lambda^2$ by a finite series in $(1-z)$ 
means that our expansions are only modified for large angular 
separations, where the renormalization terms are inherently 
ambiguous.  Our choice simply corresponds to a natural extension of 
the flat-space heat kernel which respects the periodicity of the 
two-spheres for large angular separations.  For more details, see 
Appendix~\ref{pointsplit}.

Using this procedure, one finds that to first order in the curvature  
\bea
2\sigma 
  & = &  2r^2(1-z) + \frac{r^2}{3}[1-r^2(\nabla\phi)^2](1-z)^2 
         \, , \n{4.21} \\
\Re^{\Box-\xi\,^4\!R}_0 
  & = &  1 + \frac{1}{6}\,^4\!R_{\theta\theta}(1-z) \, ,  \n{4.22} \\
\Re^{\Box-\xi\,^4\!R}_1 
  & = & \left(\frac{1}{6}-\xi\right)\,^4\!R   \, .  \n{4.23} 
\eea
Inserting these expansions into (\ref{4.8}) 
yields our approximation for the `divergent' part 
of the four-dimensional heat kernel,
\bea\n{4.24}
K_{\ind{div}}(X,X'|s) 
  & = &  \frac{1}{(4\pi s)^2}
         \exp{ \left\{ - m^2s - \frac{r^2}{2s}(1-z) \right\} }
         \left[ \vphantom{\frac{r^2}{1}} 
             1 + s \left(\frac{1}{6}-\xi\right)\,^4\!R 
             + \frac{1}{6}\,^4\!R_{\theta\theta} (1-z)
             \right. \nonumber \\
  &   &      \left. \hspace{0.8in} \mbox{} 
             - \frac{r^2}{12 s}[1-r^2(\nabla\phi)^2](1-z)^2 
         \right]  \, .
\eea  
The mode decomposition (\ref{4.18}) of this $K_{\ind{div}}$  
then boils down to evaluating the integrals  
\be\n{4.26}
J_{\ell n} \equiv -p\int_{-1}^1\!dz\,P_{\ell}(z) \,\e^{p(1-z)}\,(1-z)^n \, ,
\ee
where $p \equiv -r^2/2s$ is a dimensionless parameter and $n$ is an 
integer.  The integrals for $n\ne0$ can be obtained from the $n=0$ 
result (\ref{3.9}, \ref{3.10}) used in the flat-space case by 
differentiating with respect to $p$, yielding 
\be\n{4.27}
J_{\ell n} 
   =   \sum_{k=0}^\ell {(\ell+k)!\over k!(\ell-k)!}~{1\over (-2 p)^k}
         \left[ 
             \frac{(-1)^k 2^n}{(-2p)^n}\frac{(k+n)!}{k!}  
             -(-1)^\ell e^{2p} \sum_{\alpha=0}^n
                 \frac{2^n}{(-2p)^\alpha} 
                 \frac{(k+\alpha)!}{k!}
                 \frac{n!}{\alpha!(n-\alpha)!}
         \right]  \, .
\ee
The mode-decomposed heat kernel subtraction terms for a general 
four-dimensional spherically symmetric spacetime are then 
\be\n{4.28}
K_{\ind{div}|\ell}(x,x|s)
   =   \frac{\e^{-m^2s}}{4\pi s} \left[ 
             J_{\ell 0} + s\left(\frac{1}{6}-\xi\right)\,^4\!R\, J_{\ell 0}
             +\frac{1}{6}\,^4\!R_{\theta\theta}J_{\ell 1}
             -\frac{r^2}{12 s}[1-r^2(\nabla\phi)^2]J_{\ell 2}
         \right]  \, .
\ee
The first term in (\ref{4.28}) is the mode decomposition for flat space, 
while the other terms carry the contributions due to the curvature.
Meanwhile, the various parts of the $J_{\ell n}$ fulfill several roles. 
First, the $k\ne0$ terms in (\ref{4.27}) 
are associated with the centrifugal potential 
$\ell(\ell+1)/r^2$ induced by the the mode decomposition.  This 
potential is ignored in the renormalization in two dimensions, 
since only the first (potential-independent) 
term in the Schwinger-DeWitt expansion of 
the heat kernel contributes divergences to the two-dimensional 
Green function.  
Second, then terms in (\ref{4.27}) proportional
to $\e^{2p} = \e^{-r^2/s}$ enforce a Dirichlet boundary condition
at $r=0$, which is required if the four-dimensional subtraction term 
is to be finite there [see (\ref{4.17})].

These results are to be compared with the subtraction term in two dimensions, 
which consists of the first term of (\ref{4.13}): 
\be\n{4.29}
{\cal K}_{\ell|\ind{div}}(x,x|s) = \frac{\e^{-m^2s}}{4\pi s} \, .
\ee
In contrast to $K_{\ind{div}|\ell}$, ${\cal K}_{\ell|\ind{div}}$ 
is independent of both the position, the two-metric $h_{ab}$, and the mode 
number $\ell$.  It matches just the first term in the $k=0$ contribution 
to the flat space part of $K_{\ind{div}|\ell}$.

As an example, let us consider a quantum field in Schwarzschild space,   
for which $\,^4\!R=0$, $\,^4\!R_{\mu\nu}=0$, and  
\be\n{4.30}
[1-r^2(\nabla\phi)^2] = \frac{2M}{r} \, ,
\ee
where $M$ is the black-hole mass.  Figure~\ref{fig1} shows plots of 
$K_{\ind{div}|\ell=0}(x,x|s)$ for fixed $s$ and various values of 
$M/\sqrt{s}$.  Note that large values of $M$ cause 
the mode-decomposed subtraction terms to become negative.
\setlength{\unitlength}{1mm}
\begin{figure}[htb]
\begin{center}
\begin{picture}(80,50)
\put(-5,52){\rotatebox{-90}{\resizebox{6cm}{!}{\includegraphics{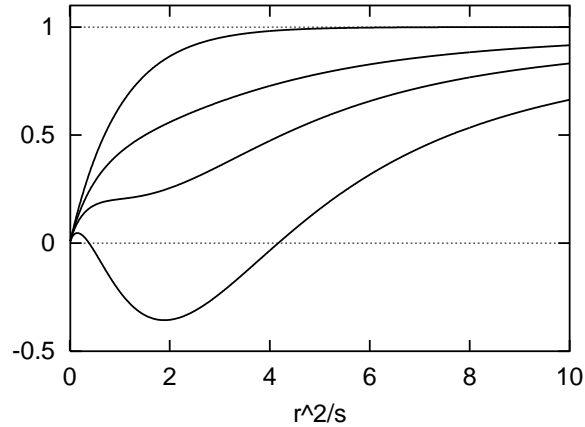}}}}
\end{picture}
\end{center}
\caption{$K_{\ind{div}|\ell=0}(x,x|s)$ in Schwarzschild space 
for fixed $s$ and $M/\sqrt{s} = 0, 2, 4, 8$ from top 
to bottom.  The common factor $\e^{-m^2s}/(4\pi s)$
has been removed. The two-dimensional subtraction term 
${\cal K}_{\ell|\ind{div}}$ would be a horizontal line at $1$ on this plot.} 
\label{fig1}
\end{figure}

The anomaly in $\langle\hat{\Phi}^2\rangle$ can now be found by integrating 
the difference of ${\cal K}_{\ell|\ind{div}}$ and $K_{\ind{div}|\ell}$ 
as in (\ref{4.19}).  We find  
\bea
\Delta\p 
  & = &  \frac{1}{2\pi}
         \left[ 
             -I[m^2s J_{\ell 0}]
             -\frac{1}{m^2}\left(\frac{1}{6}-\xi\right)\,^4\!R \,
                 I[(m^2s)^2 J_{\ell 0}]
             -\frac{1}{6}\,^4\!R_{\theta\theta} \, I[m^2s J_{\ell 1}]
             \right. \nonumber \\
  &   &      \hspace{0.5in} \left. \mbox{}
             +\frac{(mr)^2}{12}[1-r^2(\nabla\phi)^2] \, I[J_{\ell 2}]
         \right]  \, , \n{4.38}
\eea
where 
\bea\n{4.33}
I[(m^2s)^t J_{\ell n}] 
  & \equiv &  \sum_{k=2-t-n}^{\ell}{(\ell+k)! \over k!(\ell-k)!}\, 2^n \left[ 
             \frac{1}{2}\frac{(-1)^k}{(mr)^{2n+2k}}\frac{(k+n)!}{k!}(k+t+n-2)! 
         \right] 
         \nonumber \\
  &   &  \mbox{} 
         -(-1)^\ell \sum_{k=0}^{\ell}{(\ell+k)!\over k!(\ell-k)!}\, 2^n \left[
             \sum_{\alpha=0}^n \frac{(k+\alpha)!}{k!} 
             \frac{n!}{\alpha!(n-\alpha)!} 
             \frac{K_{k+t+\alpha-1}(2mr)}{(mr)^{k-t+\alpha+1}}
         \right] \,  .
\eea
(The $I[(m^2s)^t J_{\ell n}]$ result from integrating terms of the form 
$(m^2s)^{t-2} J_{\ell n}$ over $s$.)
For example, in Schwarzschild space the anomaly 
for the $\ell=0$ mode is
\be
\Delta \langle\hat{\varphi}^2_{\ell=0}\rangle
   =   \frac{1}{2\pi} K_0(2mr)
       +\frac{M}{3\pi r}\left[
           \frac{1}{(mr)^2}
           -\frac{2}{mr}K_1(2mr)
           -2K_0(2mr)
           -mr K_1(2mr) 
           \vphantom{\frac{1}{(mr)^2}}
       \right] \, . \n{4.40}
\ee         
Plots of $\Delta \langle\hat{\varphi}^2_{\ell=0}\rangle$ for 
various values of $mM$ are shown in Figure~\ref{fig2}. 
\setlength{\unitlength}{1mm}
\begin{figure}[htb]
\begin{center}
\begin{picture}(80,50)
\put(-05,52){\rotatebox{-90}{\resizebox{6cm}{!}{\includegraphics{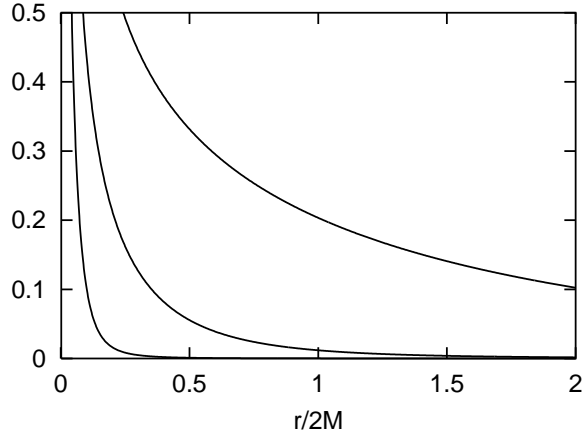}}}}
\end{picture}
\end{center}
\caption{$\Delta \langle\hat{\varphi}^2_{\ell=0}\rangle$ in Schwarzschild 
space for $mM = 0.1, 1, 10$, from top to bottom.} 
\label{fig2}
\end{figure}
Note that the anomaly in $\P$ generally diverges at any 
point $x^a$ such that $r(x^a)=0$, while for asymptotically 
flat spaces it vanishes as $r\rightarrow\infty$.


\section{The Dimensional-Reduction Anomaly in the Effective Action}
\label{W_anom}\setcounter{equation}0

In the previous section we calculated the dimensional-reduction 
anomaly in $\P$ for a general four-dimensional spherically symmetric  
space.  
We now use the same procedure to determine the 
anomaly in the effective action, denoted by 
$\Delta{\cal W}_\ell$ in (\ref{2.16}).  
Functional differentiation of $\Delta{\cal W}_\ell$ 
with respect to the metric $h_{ab}$ would then give 
the corresponding anomaly $\Delta\langle\hat{{\cal T}}_{\mu\nu}\rangle_\ell $ 
in the stress tensor in (\ref{2.17}).

For the four-dimensional effective action (\ref{4.7})  
the divergent part of the heat kernel 
consists of the first three terms of (\ref{4.8}): 
\be\label{5.1}
K_{\ind{div}}(X,X'|s) 
  =  \frac{1}{(4\pi s)^2}
     \exp{ \left\{ - m^2s - \frac{\sigma}{2s} \right\} }
     \left[ \Re^{\Box-\xi\,^4\!R}_0 + s\Re^{\Box-\xi\,^4\!R}_1 
     + s^2\Re^{\Box-\xi\,^4\!R}_2 \right]  \, .
\ee      
As in the previous section, we split the points $X$, $X'$ in the 
angular direction only.  Then one can write
\bea
2\sigma 
  & = &  2r^2 \left[ (1-z) + u(1-z)^2 + v(1-z)^3 + \cdots \right] 
         \, ,  \n{5.2} \\
\Re^{\Box-\xi\,^4\!R}_n   
  & = &  \Re^{\Box-\xi\,^4\!R}_{n(0)} + \Re^{\Box-\xi\,^4\!R}_{n(1)}(1-z) 
         + \Re^{\Box-\xi\,^4\!R}_{n(2)}(1-z)^2 + \cdots  \, , \n{5.3}
\eea
where $z=\cos\lambda$.
From the calculations for the anomaly in $\P$ we have seen that 
$u=\frac{1}{6}[1-r^2(\nabla\phi)^2]$, 
$\Re^{\Box-\xi\,^4\!R}_{0(0)}=1$, 
$\Re^{\Box-\xi\,^4\!R}_{0(1)}=\frac{1}{6}\,^4\!R_{\theta\theta}$, and 
$\Re^{\Box-\xi\,^4\!R}_{1(0)}=\left(\frac{1}{6}-\xi\right)\,^4\!R$.  
The other $\Re^{\Box-\xi\,^4\!R}_{n(k)}$ and $v$ are found in 
Appendix~\ref{pointsplit}.  Inserting these expansions into (\ref{5.1}) 
and truncating at second order in the curvature, we find
\bea\label{5.4}
K_{\ind{div}}(X,X'|s) 
 & = &  \frac{1}{(4\pi s)^2}
        \exp{ \left\{ - m^2s - \frac{r^2}{2s}(1-z) \right\} }
        \left[ \vphantom{\frac{r^1}{s^1}} 
            \left\{
                1 
                +s\Re^{\Box-\xi\,^4\!R}_{1(0)}
                +s^2\Re^{\Box-\xi\,^4\!R}_{2(0)}
            \right\}
            \right. \nonumber \\
 &   &      \mbox{}
            +\left\{
                \Re^{\Box-\xi\,^4\!R}_{0(1)}
                +s\Re^{\Box-\xi\,^4\!R}_{1(1)}
            \right\}(1-z) 
            +\left\{
                \Re^{\Box-\xi\,^4\!R}_{0(2)}
                -\frac{r^2u}{2s}
                -\frac{r^2u}{2}\Re^{\Box-\xi\,^4\!R}_{1(0)}
            \right\}(1-z)^2 
            \nonumber \\
 &   &      \left. \mbox{}
            +\left\{
                -\frac{r^2u}{2s}\Re^{\Box-\xi\,^4\!R}_{0(1)}
                -\frac{r^2v}{2s}
            \right\}(1-z)^3 
            +\frac{r^4u^2}{8s^2}(1-z)^4
        \right]  \, .
\eea    

The decomposition of the heat kernel subtraction terms (\ref{5.4}) 
is done in the same manner as in the previous section.   
Employing the definition (\ref{4.18}) of the spherical decomposition 
and using the functions $J_{\ell n}$ of (\ref{4.26}, \ref{4.27}), 
we obtain 
\bea
K_{\ind{div}|\ell}(x,x|s)
  & = &  \lim_{x'\rightarrow x} 
         2\pi r^2 \int_{-1}^1 d(\cos\lambda) \, P_\ell(\cos\lambda) \,
         K_{\ind{div}}(X,X'|s) \nonumber  \\
  & = &  \frac{\e^{-m^2 s}}{4\pi s} \left[ 
             \vphantom{\frac{r^4u^2}{8s^2}}
             \left\{ 
                 1 + s\Re^{\Box-\xi\,^4\!R}_{1(0)} +
                 s^2\Re^{\Box-\xi\,^4\!R}_{2(0)} 
             \right\}J_{\ell 0} 
             +\left\{
                 \Re^{\Box-\xi\,^4\!R}_{0(1)}+s\Re^{\Box-\xi\,^4\!R}_{1(1)}
             \right\}J_{\ell 1}
             \right. \nonumber \\
  &   &      \hspace{0.5in} \mbox{}
             +\left\{ 
                 \Re^{\Box-\xi\,^4\!R}_{0(2)} - \frac{r^2u}{2s}
                 -\frac{r^2u}{2}\Re^{\Box-\xi\,^4\!R}_{1(0)} 
             \right\}J_{\ell 2}
             \nonumber \\
  &   &      \hspace{0.5in} \left. \mbox{}
             +\left\{ 
                 -\frac{r^2v}{2s}-\frac{r^2u}{2s}\Re^{\Box-\xi\,^4\!R}_{0(1)} 
             \right\}J_{\ell 3}
             +\frac{r^4u^2}{8s^2}J_{\ell 4}
         \right]  \, .  \n{5.6}
\eea
Meanwhile, the divergences in the effective action 
for the two-dimensional theory (\ref{2.6}) arise from  
the first two terms of (\ref{4.13}):
\be\n{5.9}
{\cal K}_{\ell|\ind{div}}(x,x|s) 
   =   \frac{\e^{-m^2s}}{4\pi s} 
       \left[ 1 + s \left( \frac{1}{6}R-V_\ell \right) \right] \, .
\ee
The anomaly in the effective action is then found by integrating 
the difference of (\ref{5.6}, \ref{5.9}) as in (\ref{4.7}):  
\bea\n{5.18}
\Delta{\cal W}_\ell 
  & = &  -\frac{1}{2}\int d^2\! x \sqrt{h} \, 
         \int_0^\infty \frac{ds}{s} \left[
            {\cal K}_{\ell|\ind{div}}(x,x|s) - K_{\ind{div}|\ell}(x,x|s) 
         \right] 
         \nonumber \\ 
  & = &  \frac{m^2}{4\pi} \int d^2\! x \sqrt{h} \, \left[ 
             \vphantom{\frac{(mr)^4u^2}{8}} 
             I[J_{\ell 0}] 
             +\frac{1}{m^2} \Re^{\Box-\xi\,^4\!R}_{1(0)} 
                 I[m^2s J_{\ell 0}]
             +\frac{1}{m^4} \Re^{\Box-\xi\,^4\!R}_{2(0)} I[m^4s^2 J_{\ell 0}]
             +\Re^{\Box-\xi\,^4\!R}_{0(1)} I[J_{\ell 1}]
             \right. \nonumber \\
  &   &      \hspace{0.5in} \mbox{}
             +\frac{1}{m^2} \Re^{\Box-\xi\,^4\!R}_{1(1)} I[m^2s J_{\ell 1}]
             +(\Re^{\Box-\xi\,^4\!R}_{0(2)}-\frac{r^2u}{2}
                 \Re^{\Box-\xi\,^4\!R}_{1(0)}) I[J_{\ell 2}]             
             -\frac{(mr)^2u}{2} I[\frac{1}{m^2 s} J_{\ell 2}]
             \nonumber \\
  &   &      \hspace{0.5in} \left. \mbox{}
             -\frac{(mr)^2}{2}(u\Re^{\Box-\xi\,^4\!R}_{0(1)}+v) 
                 I[\frac{1}{m^2 s} J_{\ell 3}]
             +\frac{(mr)^4u^2}{8} I[\frac{1}{m^4 s^2} J_{\ell 4}]
         \right]  \, . \n{5.17}
\eea 
The $I[(m^2s)^t J_{\ell n}]$ are given by (\ref{4.33}).  
Using (\ref{5.18}) and the values of $u$, $v$, and the 
$\Re^{\Box-\xi\,^4\!R}_{n(k)}$ given in Appendix~\ref{pointsplit}, 
one can compute the anomalous contribution to the stress tensor.


\section{Conclusions}\label{conclusions}
\setcounter{equation}0

In a $D$-dimensional spacetime which can be foliated by 
$n$-dimensional homogeneous subspaces, 
a field can be decomposed in terms of modes on the subspaces.  
This effectively converts the system from a single quantum field 
in $D$ dimensions to a collection of fields in $(D-n)$ dimensions.
Quantities of interest for the original theory, 
such as the expectation value of the 
square of the field operator and the effective action, 
can then be written as sums of the corresponding objects 
from the dimensionally reduced theories.  
This relationship breaks down under renormalization, however, 
so that renormalized expectation values can be obtained 
by summing their lower-dimensional counterparts only if the 
contribution for each mode is modified by adding an anomalous
contribution.  This effect is the dimensional reduction anomaly.

We have explicitly calculated the anomalous contributions to 
the expectation value of the square of the field operator and 
the effective action for the case of a massive scalar field 
propagating in a general four-dimensional spherically symmetric 
space.  We have seen that the anomaly arises from several sources.  
One is the Dirichlet boundary condition imposed at $r=0$ due to 
the change in topology inherent in the spherical reduction of the 
spacetime.
Other contributions are more local in nature, arising from the 
dimension-dependent contributions of the curvature and field potential 
to divergences.
The resulting anomaly terms are constructed from the curvature, 
the dilaton field, and their covariant derivatives, and cannot be 
eliminated by further finite renormalization.  

The anomalies calculated in this paper may be of importance to 
recent attempts to calculate the stress tensor and Hawking radiation 
in black-hole spacetimes using quantum fields in two dimensions 
\cite{MuWiZe:94,NoOd:99}.  These attempts are based on the dimensional 
reduction of a massless minimally-coupled quantum field in a 
Schwarzschild spacetime, followed by renormalization in two dimensions.
We have seen, however, that the contributions of the dimensionally 
reduced fields should be modified by adding the corresponding 
anomaly term.  We intend to return to a discussion of this 
interesting topic in a future publication.



\bigskip

\vspace{12pt}
{\bf Acknowledgments}:\ \  
The author would like to thank Valeri Frolov and Andrei Zelnikov 
for useful discussions.
This work was supported by the
Natural Sciences and Engineering Research Council of Canada. 

\bigskip


\appendix

\section{Spherical Decomposition of Curvatures}
\label{sphdecomp}
\setcounter{equation}0

Consider a line element of the form
\be
ds^2  =  g_{\mu\nu} dX^\mu dX^\nu
      =  h_{ab}dx^a dx^b + \rho^2 \e^{-2\phi} \omega_{ij}dy^i dy^j ~,
\ee
where $h_{ab} = h_{ab}(x^c)$ is an arbitrary two-dimensional metric and
$\omega_{ij} = \omega_{ij}(y^k)$ is the metric of a two-sphere.  
The dilaton $\phi$ is a function of the $x^a$ only, and $\rho$ is a constant
with the dimensions of length.  
The radius of a two-sphere of fixed $x^a$ is 
$r=\rho\e^{-\phi}$.

We wish to decompose our field theory in terms of
modes on the two-sphere.  This requires rewriting four-dimensional geometric 
quantities like the curvatures in terms of the corresponding curvatures 
for the metric $h$.

Our notational conventions are as follows: four-dimensional covariant 
derivatives are denoted by $()_{;a}$, while $\Box$ is understood to 
represent the d'Alembertian with respect to $g$.  
Meanwhile, $\nabla$, $()_{|a}$ and $\Delta$ are the two-dimensional 
covariant derivatives and d'Alembertian calculated using the metric $h_{ab}$.
For the dilaton $\phi$ we shall understand $\phi_a$, $\phi_{ab}$, etc. to
denote multiple two-dimensional covariant derivatives of $\phi$.
For example, the four-dimensional d'Alembertian of 
an angle-independent scalar $S$ decomposes to 
\be
\Box S = \Delta S - 2\nabla\phi\!\cdot\!\nabla S \, .
\ee
In particular, 
\be
\Box \phi  =  \Delta \phi - 2(\nabla\phi)^2 \, . 
\ee

For the given line element, the nonvanishing Christoffel symbols are
\bea
\,^4\Gamma^a_{bc}[g]  & = &  \,^2\Gamma^a_{bc}[h]  \, ,   \\
\,^4\Gamma^a_{ij}[g]  & = &  \phi^a g_{ij}     \, ,   \\
\,^4\Gamma^i_{ja}[g]  & = &  -\phi_a \delta^i_j  \, , \\
\,^4\Gamma^k_{ij}[g]  & = &  \,^{2}\Gamma^k_{ij}[\omega]     \, .
\eea
Selecting coordinates $(\theta,\eta)$ on the two-spheres,
where 
\be
d\omega_{ij}y^iy^j = d\theta^2 + \sin^2\!\theta\, d\eta^2 \, ,
\ee
one finds 
\be
\,^4\Gamma^\theta_{\eta\eta}[g]  =  -\sin\theta\cos\theta  
\, , \qquad 
\,^4\Gamma^\eta_{\eta\theta}[g]  =  \frac{\cos\theta}{\sin\theta}
\, .
\ee

For convenience, we define the following commonly-occurring functions 
of the dilaton field:
\bea
A & = & 1-r^2(\nabla\phi)^2     \, ,      \\
B & = & \Delta\phi-2(\nabla\phi)^2   \, , \\
T_{ab} & = & \phi_{ab} - \phi_a\phi_b  \, , \\
T & = & h^{ab}T_{ab} \, = \, \Delta\phi-(\nabla\phi)^2 \, .
\eea
Since the two-sphere metric has 
constant curvature $\,^2\!R[\omega] = 2$, explicit reference
to it may be dropped.  Henceforth we shall assume all curvatures
to be with respect to the two-dimensional metric $h_{ab}$ unless explicitly
labelled otherwise.  Using this notation, one can show that 
the only nonvanishing components of the four-dimensional curvatures 
are  
\bea
\,^4\!R_{abcd}[g]  & = &  \frac{1}{2}\,R\,(h_{ac}h_{bd}-h_{ad}h_{bc})\, , \\
\,^4\!R_{aibj}[g]  & = &  g_{ij} T_{ab}   \, , \\
\,^4\!R_{ijkm}[g]  & = &  \frac{A}{r^2} (g_{ik}g_{jm}-g_{im}g_{jk})\, , \\
                   &   &  \nonumber \\
\,^4\!R_{ab}[g]    & = &  \frac{1}{2}\,R\,h_{ab}+2T_{ab} \, ,   \\
\,^4\!R_{ij}[g]    & = &  g_{ij}\left[ \frac{1}{r^2}+B \right]\, , \\
                   &   &  \nonumber \\
\,^4\!R[g]         & = &  R+4\Delta\phi-6(\nabla\phi)^2+\frac{2}{r^2} 
                          \, ,
\eea
while the only nonvanishing $\,^4\!R_{\alpha\beta;\gamma}$ are 
\bea
\,^4\!R_{ab;c}[g]  
  & = &  \frac{1}{2}h_{ab}\,R_{|c} + 2T_{ab|c} \, ,   \\
\,^4\!R_{am;n}[g]  
  & = &  g_{mn}\left[ 
             \left(-\frac{1}{2}\,R+\frac{1}{r^2}+B\right)\phi_a 
             -2 T_{ab}\phi^b  
         \right]  \, ,   \\
\,^4\!R_{mn;a}[g]  
  & = &  g_{mn} \left(\frac{1}{r^2} + B \right)_{|a}    \, .
\eea
Also, 
\bea
\,^4\!R_{mn;ab}[g]
  & = &  g_{mn} \left(\frac{1}{r^2} + B \right)_{|ab} \, , \\
\,^4\!R_{mn;jk}[g]
  & = &  -\left( g_{km} g_{nj} + g_{kn} g_{mj} \right) \left[
           \left(-\frac{1}{2}\,R+\frac{1}{r^2}+B\right)(\nabla\phi)^2
           -2 T_{ab}\phi^a\phi^b
         \right]  
         \nonumber \\ 
  &   &  \mbox{}
         -g_{jk} g_{mn}\left(\frac{1}{r^2} + B \right)_{|a} \phi^a \, ,  \\ 
\Box \,^4\!R_{mn}[g]
  & = &  g_{mn} \left\{ 
             \left[\Delta-2\nabla\phi\cdot\nabla\right]
                 \left(\frac{1}{r^2}+B\right)
             +R(\nabla\phi)^2
             -2\left(\frac{1}{r^2} + B \right)(\nabla\phi)^2
             \right. \nonumber \\
  &   &      \left. \hspace{0.25in} \mbox{}
             +4T_{ab}\phi^a\phi^b  
         \right\} \, , \\ 
  &   &  \nonumber \\
\,^4\!R_{;a} 
  & = &  \left(R + 4\Delta\phi - 6(\nabla\phi)^2 
         + \frac{2}{r^2} \right)_{|a} \, ,  \\
  &   &  \nonumber \\
\,^4\!R_{;ab} 
  & = &  \left(R + 4\Delta\phi - 6(\nabla\phi)^2 
         + \frac{2}{r^2} \right)_{|ab}  \, , \\
\,^4\!R_{;mn} 
  & = &  -g_{mn} \left( R + 4\Delta\phi - 6(\nabla\phi)^2 
         + \frac{2}{r^2} \right)_{|a} \phi^a \, ,  \\
\Box\,^4\!R
  & = &  [\Delta-2\nabla\phi\cdot\nabla]
             \left( R + 4\Delta\phi - 6(\nabla\phi)^2
                 + \frac{2}{r^2} \right)    \, .
\eea

\section{Point Splitting}
\label{pointsplit}
\setcounter{equation}0

It will be necessary to write short-distance expansions for
$\sigma$ and the $D^{\frac{1}{2}}a_n$ for $X$ and $X'$ separated
along the two-spheres.  We follow a method similar to that developed
in \cite{FrSuZe:00}.  Without loss of generality we take the points 
to be split in the $\theta$ direction only, with angular separation
$\lambda = \theta - \theta'$.  Our procedure will be to calculate 
the desired quantities first as expansions in powers of $\lambda^2$,
and then to convert them to expansions in powers of $(1-\cos\lambda)$ for
use in the mode-decomposition calculations.

We take as our ansatz for the geodetic interval $\sigma$
\be\n{B1}
2\sigma(x,y;x',y')
  = (\tilde{r}\lambda)^2 + U(\tilde{x})(\tilde{r}\lambda)^4
    + V(\tilde{x})(\tilde{r}\lambda)^6 + \cdots  \, ,
\ee 
where $\tilde{x} \equiv \frac{1}{2}(x+x')$.  Taking the derivative
of $\tilde{\sigma}$ with respect to each of the coordinates and 
requiring $\sigma=\frac{1}{2}g^{\alpha\beta}\sigma_\alpha\sigma_\beta$
in the coincidence limit, one can show that  
\bea
U(x) & = & -\frac{1}{12}(\nabla\phi)^2 \, , \n{B2} \\
V(x) & = & \frac{1}{90}(\nabla\phi)^4 
           -\frac{1}{120}\phi^a\phi^b\phi_{ab} \, , \n{B3} 
\eea 
and 
\bea
(\sigma^\theta)^2 
  & = &  \lambda^2 \left[ 
             1 - \frac{1}{3}r^2(\nabla\phi)^2\lambda^2 
             +r^4\left( \frac{17}{180}(\nabla\phi)^4
                 -\frac{1}{20}\phi^a\phi^b\phi_{ab} \right)\lambda^4
             +\cdots
         \right]  \, ,  \n{B4} \\
\sigma^\eta
  & = &  0  \, , \n{B5} \\
\sigma^a
  & = &  -\frac{1}{2}\phi^a(r\lambda)^2 
         +\left[ -\frac{1}{24}\phi^{ab}\phi_b 
             +\frac{1}{12}(\nabla\phi)^2\phi^a \right](r\lambda)^4 
         + \cdots \, . \n{B6}
\eea

The expansion (\ref{B1}) for $\sigma$ can be converted into one
in terms of $(1-\cos\lambda)$ using  
\be\n{B7}
\lambda^2 = 2(1-z)+\frac{1}{3}(1-z)^2+\frac{4}{45}(1-z)^3 + \cdots \, ,
\ee
where $z\equiv\cos\lambda$.
Defining the functions $u(x)$, $v(x)$ by
\be
2\sigma(x,y;x,y')
  =  2r^2\left[ (1-z) + u(x)(1-z)^2 + v(x)(1-z)^3 + \cdots \right] \, ,
\ee  
we obtain
\bea
u(x) & = & \frac{1}{6}[1-r^2(\nabla\phi)^2] \, ,  \\
v(x) & = & \frac{2}{45}\left[
               1-\frac{5}{4}r^2(\nabla\phi)^2+r^4(\nabla\phi)^4
               -\frac{3}{8}r^4\nabla\phi\cdot\nabla[(\nabla\phi)^2]
           \right] \, .
\eea

Combining (\ref{B4}--\ref{B7}) with the results of
Appendix~\ref{sphdecomp} and the 
short-distance expansions of \cite{Chri:76,Chri:78}, one can derive
expansions for the $D^{\frac{1}{2}}a_n$ in powers of $(1-z)$.  
Writing 
\be
D^{\frac{1}{2}}a^{\Box-\xi\,^4\!R}_n 
  =  \Re^{\Box-\xi\,^4\!R}_n 
  =  \Re^{\Box-\xi\,^4\!R}_{n(0)} 
     +\Re^{\Box-\xi\,^4\!R}_{n(1)}(1-z) 
     +\Re^{\Box-\xi\,^4\!R}_{n(2)}(1-z)^2 
     +\cdots  ~,
\ee
one can show that
\bea
\Re^{\Box-\xi\,^4\!R}_{0(0)}
  & = &  1  
\, , \\
\Re^{\Box-\xi\,^4\!R}_{0(1)}
  & = &  \frac{1}{6}(1+r^2B) 
\, , \\
\Re^{\Box-\xi\,^4\!R}_{0(2)}
  & = &  \frac{1}{90}A^2 
         +\frac{1}{72}(1+r^2B)^2 
         +\frac{1}{36}(1+r^2B)(1-4r^2(\nabla\phi)^2)
         \nonumber \\
  &   &  \mbox{}
         +\frac{r^4}{180}\left[
             \vphantom{\left(\frac{1}{r^2}\right)_{\!|a}}
             \frac{3}{2}R(\nabla\phi)^2 
             +6T_{ab}\phi^a\phi^b
             +2T_{ab}T^{ab}
             +12\left(\frac{1}{r^2}+B\right)(\nabla\phi)^2
             \right. \nonumber \\
  &   &      \left. \hspace{0.5in} \mbox{}
             +6\left(\frac{1}{r^2}+B\right)_{\!|a}\!\phi^a
         \right] 
\, , \\
 & & \nonumber \\
\Re^{\Box-\xi\,^4\!R}_{1(0)}
  & = &  \left(\frac{1}{6}-\xi\right) \left(
             R+4\Delta\phi-6(\nabla\phi)^2+\frac{2}{r^2} 
         \right)
\, , \\
\Re^{\Box-\xi\,^4\!R}_{1(1)}
  & = &  \frac{1}{6}\left(\frac{1}{6}-\xi\right)
             \left[(1+r^2B)+r^2\nabla\phi\cdot\nabla\right] 
             \left(
                 R+4\Delta\phi-6(\nabla\phi)^2+\frac{2}{r^2}
             \right) 
             \nonumber \\
  &   &  \mbox{}
         +\frac{r^2}{180}\left[
             RT 
             +3 R(\nabla\phi)^2
             +8 T_{ab}T^{ab} 
             +12 T_{ab}\phi^a\phi^b
             +\left(
                 R+4\Delta\phi-6(\nabla\phi)^2+\frac{2}{r^2}
             \right)_{\!|a}\!\phi^a
             \right. \nonumber \\
  &   &      \left. \hspace{0.5in} 
             \vphantom{\left[\frac{2}{r^2}\right]_{|a}}  
             \mbox{}
             + 3[\Delta-2\nabla\phi\cdot\nabla]\left(\frac{1}{r^2}+B\right)
             - 6\left(\frac{1}{r^2}+B\right)(\nabla\phi)^2
         \right] 
         \nonumber \\
  &   &  \mbox{}
         +\frac{1}{90r^2}\left(
             2A^2 + A(1+r^2B) - 2(1+r^2B)^2
         \right)
\, , \\       
 & & \nonumber \\
\Re^{\Box-\xi\,^4\!R}_{2(0)}
  & = &  \frac{1}{2}\left(\frac{1}{6}-\xi\right)^2
             \left(
                 R+4\Delta\phi-6(\nabla\phi)^2+\frac{2}{r^2}
             \right)^2 
         \nonumber \\
  &   &  \mbox{}
         +\frac{1}{6}\left(\frac{1}{6}-\xi\right)
             \left[ \Delta-2\nabla\phi\cdot\nabla \right]
             \left(
                 R+4\Delta\phi-6(\nabla\phi)^2+\frac{2}{r^2}
             \right) 
         \nonumber \\
  &   &  \mbox{}
         +\frac{1}{180}\left[
             \left[\Delta-2\nabla\phi\cdot\nabla \right]
             \left(
                 R+4\Delta\phi-6(\nabla\phi)^2+\frac{2}{r^2}
             \right) 
             \right. \nonumber \\
  &   &      \left. \hspace{0.5in} \mbox{}
             + \frac{1}{2}R^2 - 2RT
             + 4T_{ab}T^{ab} + \frac{4}{r^4}A^2 
             - \frac{2}{r^4}(1+r^2B)^2
         \right] \, .
\eea
It is easily verified that for flat spacetime each of the 
$\Re^{\Box-\xi\,^4\!R}_{n(k)}$ vanishes, 
except for $\Re^{\Box-\xi\,^4\!R}_{0(0)}$.



\end{document}